\documentclass[aps,preprint,preprintnumbers,amsmath,amssymb]{revtex4}
\usepackage{amsmath,mathrsfs,amsbsy,color,graphicx,bm,amsthm,amsfonts}
\usepackage{units}
\usepackage{bbm}
\usepackage{times}
\usepackage{dcolumn}
\usepackage{mathrsfs}
\usepackage{amsmath,amssymb,epsfig}

\newcommand{\be}{\begin{equation}}
\newcommand{\ee}{\end{equation}}
\newcommand{\bea}{\begin{eqnarray}}
\newcommand{\eea}{\end{eqnarray}}

\renewcommand{\det}{{\rm Det}\,}
\newcommand{\gr}[1]{\boldsymbol{#1}}

\newcommand{\N}{{\cal N}}

\newcommand{\sig}{\gr{\sigma}}

\newcommand{\eq}[1]{Eq.~(\ref{#1})}

\begin{document}

\title{Generation of  genuine tripartite entanglement  for continuous variables  in de Sitter space}
\author{ Jieci Wang, Cuihong Wen\footnote{Email: cuihong\_wen@hnu.edu.cn}, Songbai Chen\footnote{Email: csb3752@hunnu.edu.cn}, and Jiliang Jing\footnote{Email: jljing@hunnu.edu.cn}}
\affiliation{Department of Physics, and Collaborative Innovation Center for Quantum Effects \\
and Applications,
 Hunan Normal University, Changsha, Hunan 410081, China
}


\begin{abstract}
We study the distribution of quantum entanglement  for continuous variables among  causally disconnected open charts in de Sitter space. It is found that  genuine tripartite entanglement is generated among the  open chart modes under the influence of curvature of de Sitter space for any nonzero squeezing.  Bipartite entanglement is also generated when the curvature is strong enough,  even though the observers are separated by the event horizon. This provides a clearcut interpretation of the two-mode squeezing mechanism in the de Sitter space. In addition, the curvature generated genuine tripartite entanglement  is found to be less sensitive to the mass parameter than the generated bipartite entanglement.  The effects of the curvature of de Sitter space on the generated entanglement become more apparent  in the limit of  conformal and massless scalar fields.

\end{abstract}

\vspace*{0.5cm}

\maketitle
\section{Introduction}

Quantum entanglement, predicted by  Schr\"odinger  in 1935 \cite{schr}, has fascinated many physicists since it was put forward due to its highly counterintuitive properties.
As a central concept in quantum information theory,  entanglement
represents nonlocal correlation in quantum systems \cite{Bell}.   Entanglement
is needed as key resource to carry out some quantum information processing tasks, e.g.,
quantum teleportation,  quantum computation, quantum simulation, quantum superdense coding, quantum error
correction, and so on \cite{Nielsen}.  It was recently realized that understanding the entanglement between field modes
near a horizon, either  black hole horizon or cosmic horizon, can help to
understand some  key questions in black hole thermodynamics
and their relation to information \cite{Bombelli-Callen,birelli, Hawking-Terashima,Brout}. Therefore, many efforts
have been expended on the investigation of  entanglement in
relativistic quantum systems \cite{Peres, Schuller-Mann, RQI1,jieci1, jieci2, Ralph,RQI6,
adesso2,RQI2,RQI4,RQI5,RQI7, adesso3}, and  in the context of cosmology \cite{Ball:2005xa,Fuentes:2010dt,Nambu:2011ae, Kanno16, wang2015}.
In particular,  we know that any two mutually separated regions would become causally disconnected in the exponentially expanding de Sitter space \cite{Sasaki:1994yt}.  The Bogoliubov transformations
between  the open chart vacua and the  Bunch-Davies vacuum  which  have support on both regions of a free massive scalar field were derived in~\cite{Sasaki:1994yt}. Later, quantum entanglement and related  nonlocal  correlations of free  field modes were studied in de Sitter space \cite{Maldacena:2012xp,Albrecht18, Kanno:2014lma,Kanno:2016qcc,Choudhury:2017bou,Iizuka:2014rua, Nambu2018,Kanno:2014ifa, Dimitrakopoulos:2015yva}.

In this paper we study  the distribution of   quantum entanglement  for continuous variables  in de Sitter space.
 We consider the sharing of entanglement among three subsystems: the subsystem $A$ observed by a global observer Alice, the subsystem  $B$ observed by Bob in  region $R$ of the de Sitter space, and the subsystem $\bar B$ for an imaginary observer anti-Bob in  region $L$.
The initial state between modes $A$ and $B$ is prepared in a two-mode squeezed state from the perspective of  a global observer. Our studying on the behaviors of continuous variables entanglement  in  de Sitter space
is motivated by the following two reasons. Firstly,  the two-mode squeezed state can be used  to define particle states when the spacetime has at least two
asymptotically flat regions \cite{birelli,Peres,Schuller-Mann,Ralph,
adesso2,RQI2,RQI4,RQI5, adesso3}.  This  state has a special role in quantum field
theory because the  field modes in causally disconnected regions  are found to be pair-wise squeezed
due to spacetime  evolution and relativistic effects \cite{birelli,Peres,Schuller-Mann,Ralph,
adesso2,RQI2,RQI4,RQI5, adesso3}. Secondly, as a paradigmatic entangled state for continuous variables,  the  two-mode squeezed state can
be produced in the lab and exploited for many current realization of continuous variable
 quantum information tasks \cite{brareview}.

This work aims to study how the space curvature of  de Sitter space  affects the
continuous variable  entanglement described by observers in open chart vacua.  We will  derive the phase-space description of quantum state evolution in the de Sitter space   basing on the Bogoliubov transformation between different open chart vacua and the Bunch-Davies vacuum.  Then we evaluate not only the initial bipartite entanglement as degraded by the  expanding de Sitter space, but
remarkably, the multipartite entanglement which arises among all
 open chart modes. We are going to demonstrate that the bipartite entanglement initially prepared in the global  bipartite modes does not disappear, but is redistributed  among the modes in different open charts.  As a consequence of
the monogamy of entanglement, the entanglement between the initial modes
described by the globe  observers is degraded. In addition,  the modes observed by Bob and anti-Bob are entangled when the curvature is strong enough  even though they are separated by the event horizon, verifying the  nonlocal nature of entanglement in curved space.

The outline of the paper is as follows. In Sec. II we review the solutions of mode functions and Bogoliubov transformations  in the de Sitter space.  In Sec. III we  discuss the measurements of bipartite and tripartite Gaussian quantum entanglement. In Sec. IV we study the distribution of Gaussian quantum entanglement and the behaviors of the generated entanglement under the influences of the expanding and  curvature of  de Sitter space. The last section is devoted to a brief summary.

\section{Quantization of  scalar field in de Sitter space \label{model}}

We consider a free scalar field $\phi$ with mass $m$ initially prepared in the Bunch-Davies vacuum of de Sitter space with metric $g_{\mu\nu}$.
 The action of the field is given by
\begin{eqnarray}
S=\int d^4 x\sqrt{-g}\left[\,-\frac{1}{2}\,g^{\mu\nu}
\partial_\mu\phi\,\partial_\nu \phi
-\frac{m^2}{2}\phi^2\,\right]\,.
\end{eqnarray}
Then we assume  that one subsystem of the initial state is described  the experimenter  Bob who  stays in the open region $R$ of the de Sitter space.
  The coordinate frames of open charts in de Sitter space  can be obtained by analytic continuation from the Euclidean metric, and  region $R$  is causally disconnected from region $L$ \cite{Sasaki:1994yt}.  The metrics for the  open charts $R$ and $L$ in the de Sitter space
are given by
\begin{eqnarray}
ds^2_R&=&H^{-2}\left[-dt^2_R+\sinh^2t_R\left(dr^2_R+\sinh^2r_R\,d\Omega^2\right)
\right]\,,\nonumber\\
ds^2_L&=&H^{-2}\left[-dt^2_L+\sinh^2t_L\left(dr^2_L+\sinh^2r_L\,d\Omega^2\right)
\right]\,,
\end{eqnarray}
where  $H^{-1}$ is the Hubble radius and $d\Omega^2$ is the metric on the two-sphere.

Solving the Klein-Gordon equation for the scalar field $\phi$  in different open regions, one obtains
\begin{eqnarray}\label{solutions1}
u_{\sigma p\ell m}(t_{R(L)},r_{R(L)},\Omega)&\sim&\frac{H}{\sinh t_{R(L)}}\,
\chi_{p,\sigma}(t_{R(L)})\,Y_{p\ell m} (r_{R(L)},\Omega)\,,\qquad \nonumber\\
-{\rm\bf L^2}Y_{p\ell m}&=&\left(1+p^2\right)Y_{p\ell m}\,,
\end{eqnarray}
where $Y_{p\ell m}$ are harmonic functions on the three-dimensional hyperbolic space. In Eq. (\ref{solutions1}) $\chi_{p,\sigma}(t_{R(L)})$ are positive frequency mode functions supporting  on the $R$ and $L$ regions ~\cite{Sasaki:1994yt}
\begin{eqnarray}
\chi_{p,\sigma}(t_{R(L)})=\left\{
\begin{array}{l}
\frac{e^{\pi p}-i\sigma e^{-i\pi\nu}}{\Gamma(\nu+ip+\frac{1}{2})}P_{\nu-\frac{1}{2}}^{ip}(\cosh t_R)
-\frac{e^{-\pi p}-i\sigma e^{-i\pi\nu}}{\Gamma(\nu-ip+\frac{1}{2})}P_{\nu-\frac{1}{2}}^{-ip}(\cosh t_R)
\,,\\
\\
\frac{\sigma e^{\pi p}-i\,e^{-i\pi\nu}}{\Gamma(\nu+ip+\frac{1}{2})}P_{\nu-\frac{1}{2}}^{ip}(\cosh t_L)
-\frac{\sigma e^{-\pi p}-i\,e^{-i\pi\nu}}{\Gamma(\nu-ip+\frac{1}{2})}P_{\nu-\frac{1}{2}}^{-ip}(\cosh t_L)
\,,
\label{solutions}
\end{array}
\right.
\end{eqnarray}
where $P^{\pm ip}_{\nu-\frac{1}{2}}$ are the associated Legendre functions
and  $\sigma=\pm 1$ is employed to  distinguish  the
independent solutions in each open region.  In addition, $p$ is a positive real parameter normalized by $H$,  and $\nu$ is a mass parameter
$
\nu=\sqrt{\frac{9}{4}-\frac{m^2}{H^2}}\,
$. These solutions can be normalized by the  factor
$
N_{p}=\frac{4\sinh\pi p\,\sqrt{\cosh\pi p-\sigma\sin\pi\nu}}{\sqrt{\pi}\,|\Gamma(\nu+ip+\frac{1}{2})|}\,.
$
To make the discussion clear we should know that  $p$ can be regarded as the curvature parameter of the de Sitter space because, as $p$ becomes smaller than $1$, the effect of  curvature gets stronger and stronger \cite{Kanno16, Albrecht18}.
On the other hand , the mass parameter  $\nu$ has two special values, which are
 $\nu=1/2$  for the conformally coupled
massless scalar field  and $\nu=3/2$ for the minimally coupled massless limit.

The scalar field can be
 expanded in terms of the creation and annihilation operators:
\begin{eqnarray}
\hat\phi(t,r,\Omega)
=\frac{H}{\sinh t}\int dp \sum_{\sigma,\ell,m}\left[\,a_{\sigma p\ell m}\,\chi_{p,\sigma}(t)
+a_{\sigma p\ell -m}^\dagger\,\chi^*_{p,\sigma}(t)\right]Y_{p\ell m}(r,\Omega)
\,,
\end{eqnarray}
where
$a_{\sigma p\ell m}|0\rangle_{\rm BD}=0$ is the  annihilation operator of the Bunch-Davies vacuum.
For simplicity, hereafter we omit the indices $p$, $\ell$, $m$ in the  operators $\phi_{p\ell m}$,
$a_{\sigma p\ell m}$ and $a_{\sigma p\ell -m}^\dag$.
Similarly, the mode functions and the associated Legendre
functions  are rewritten as
 $\chi_{p,\sigma}(t)\rightarrow\chi^{\sigma}$,
$P_{\nu-1/2}^{ip}(\cosh t_{R,L})\rightarrow P^{R, L}$, and
$P_{\nu-1/2}^{-ip}(\cosh t_{R,L})\rightarrow P^{R*, L*}$.

On the other hand one can consider the positive frequency
mode functions \cite{Sasaki:1994yt, Kanno16, Albrecht18}
\begin{eqnarray}
\varphi^q=\left\{
\begin{array}{ll}
\frac{|\Gamma(1+ip)|}{\sqrt{2p}}P^q~&\mbox{in region}~q\,,
\\
0~ &\mbox{in the opposite region}\,,
\end{array}
\right.
\label{varphi}
\end{eqnarray}
which
 are defined only on the $q=(R, L)$ region, respectively.  In this case we introduce the creation and annihilation operators ($b_q,b_q^\dag$) in different regions by
 $b_q|0\rangle_{q}=0$. Then we can relate the creation and annihilation operators $(a_\sigma,a_\sigma^\dag)$ and $(b_q,b_q^\dag)$ in different reference frame
 by the Bogoliubov transformation
\begin{eqnarray}
\phi(t)=a_\sigma\,\chi^\sigma+a_\sigma^\dag\,\chi^\sigma{}^*
=b_q\,\varphi^q+b_q^\dag\,\varphi^q{}^*\,.
\label{fo}
\end{eqnarray}

Using this  transformation, the Bunch-Davies vacuum can be constructed from the vacuum states
over $|0\rangle_{q}$ in the regions $R$ and $L$  \cite{Sasaki:1994yt, Kanno16, Albrecht18}, which is
\begin{eqnarray}
|0\rangle_{\rm BD} = N_{\gamma_p}^{-1}
\exp\left(\gamma_p\,c_R^\dagger\,c_L^\dagger\,\right)|0\rangle_{R}|0\rangle_{L}\,,
\label{bogoliubov3}
\end{eqnarray}
where the parameter $\gamma_p$ is given by
\begin{eqnarray}
\gamma_p = i\frac{\sqrt{2}}{\sqrt{\cosh 2\pi p + \cos 2\pi \nu}
 + \sqrt{\cosh 2\pi p + \cos 2\pi \nu +2 }}\,.
\label{gammap2}
\end{eqnarray}

The normalization factor $N_{\gamma_p}$ in Eq.~(\ref{bogoliubov3})  is found to be \cite{Sasaki:1994yt, Kanno16, Albrecht18}
\begin{eqnarray}
N_{\gamma_p}^2
=\left|\exp\left(\gamma_p\,c_R^\dagger\,c_L^\dagger\,\right)|0\rangle_{R}|0\rangle_{L}
\right|^2
=\frac{1}{1-|\gamma_p|^2}\,.
\label{norm2}
\end{eqnarray}
In addition,  $\gamma_p$ simplifies to $|\gamma_p|=e^{-\pi p}$ for the conformally coupled
massless scalar ($\nu=1/2$) and the minimally coupled massless scalar ($\nu=3/2$). It is worthy to note that the limit of large p has small  $\gamma_p$  while small $p$ (i.e. larger curvature) has $\gamma_p$ approaching 1 ($\nu=1/2$). This because the effect of  curvature becomes stronger and stronger when $p$ becomes smaller and smaller.

\section{Quantifying entanglement for continuous variables \label{Gentanglement}}
In this section we recall the measures of Gaussian quantum entanglement for continuous variables.
The character of a  Gaussian state can be  completely described in phase space
by the symmetric  covariance matrix
 $\gr{\sigma}$, whose entries are
$\sigma_{ij}=\text{Tr}\big[ {{{\{ {{{\hat R}_i},{{\hat R}_j}} \}}_ + }\ {\rho _{AB}}} \big]$. Here
$\hat{R}= (\hat x_1^A,\hat p_1^A, \ldots ,\hat x_n^A,\hat p_n^A,\hat x_1^B,\hat p_1^B, \ldots ,\hat x_m^B,\hat p_m^B)^{\sf T}$ is a
vector  grouped the quadrature  field  operators. For each mode $i$,  the
phase space variables are defined  by $\hat a_i^A=\frac{\hat x_i^A+i\hat p_i^A}{\sqrt{2}}$ and $\hat a_i^B=\frac{\hat x_i^B+i\hat p_i^B}{\sqrt{2}}$, where $\hat a_i^A$ and $\hat a_i^B$ are annihilation operators of the subsystems. The   canonical commutation relations of these operators can be expressed as $[{{{\hat R}_i},{{\hat R}_j}} ] = i{\Omega _{ij}}$, with the symplectic form $\Omega  =  \bigoplus _1^{n+m} {{\ 0\ \ 1}\choose{-1\ 0}}$.
The covariance matrix $\sigma_{AB}$ must satisfy  the
Robertson-Schr\"odinger uncertainty relation \cite{simon87}
\begin{equation}\label{bonfide}
\gr{\sigma}+i\Omega \geq 0\,,
\end{equation}
to describe a physical state.

To  measure bipartite entanglement in a relativistic setting,  we employ the contangle \cite{contangle}
 which is an entanglement monotone under
Gaussian local operations and classical communication. This choice is because
our main focus is the effect of spacetime  curvature of de Sitter space on the distribution of entanglement among field modes
in different open charts. In this setting, the
contangle is the measure which enables  mathematical treatment of
distributed continuous variables entanglement as emerging from the fundamental
monogamy constraints \cite{contangle,hiroshima,pisa}.
For pure states the contangle
$\tau$ is defined  as the square of the logarithmic
negativity and it can be  extended to mixed states by the Gaussian
convex roof  \cite{ordering}. If $\sig_{AB} $ is
the covariance matrix of a mixed bipartite Gaussian state where
subsystem $A$ comprises one mode only,  the contangle $\tau$ can be computed by  (for  detail please see Appendix A) \cite{contangle}
\begin{equation}
\label{tau} \tau (\sig_{AB} )\equiv \tau (\sig_{A\vert
B}^{opt} )=g[m_{AB}^2 ],\;\;\;g[x]={\rm arcsinh}^2[\sqrt {x-1}],
\end{equation}
where $\sig_{AB}^{opt} $ corresponds to a pure Gaussian state,
and $m_{AB} \equiv m(\sig_{AB}^{opt} )=\sqrt {\det
\sig_A^{opt} } =\sqrt {\det \sig_B^{opt} } $.
 The reduced covariance matrix $\sig_{A(B)}^{opt} $ of subsystem $A (B)$ are obtained by
tracing over the degrees of freedom of subsystem $B$ ($A)$.

 Unlike classical correlations,
entanglement is  monogamous, which means that it { \it can not} be freely
shared among multiple subsystems of a mulitpartite quantum system
\cite{pisa}. Because of this property, we can employ the residual multipartite entanglement as a measurement of nonclassical correlations by exploring the entanglement distributed
between multipartite systems. For a state distributed among $N$ parties each
owns a single qubit (or a single mode), the monogamy
constraint is described by the Coffman-Kundu-Wootters inequality
\cite{ckw},
\begin{equation}
\label{ckwine} E_{S_i \vert (S_1 \ldots S_{i-1} S_{i+1} \ldots S_N
)} \ge \sum\limits_{j\ne i}^N {E_{S_i \vert S_j } },
\end{equation}
where the  multipartite system is partitioned in $N$ subsystems $S_k$
($k=1,{\ldots},N$), each owning a single mode, and $E$ is the
measure of bipartite entanglement between the $i$-th and $j$-th subsystems. The left hand side of
inequality (\ref{ckwine}) measures the bipartite entanglement
between a probe subsystem $S_i $ and the whole $N-1$ remaining subsystems.  The other side  quantifies the total bipartite
entanglement between $S_i$ and each one of the remaining subsystems
$S_{j\ne i}$ in the reduced  subsystems. The
residual multipartite entanglement is defined by the non-negative
difference between these two entanglements, minimized over all
choices of the probe subsystem. 

In the simplest case of tripartite quantum system, the residual
entanglement has the meaning of the genuine tripartite entanglement
shared by the three subsystems \cite{ckw}.  In this case the multipartite entanglement  is measured by
\cite{contangle}
\begin{equation}\label{taures}
\tau(\sig_{i|j|k})\equiv\min_{(i,j,k)} \left[
\tau(\sig_{i|(jk)})-\tau(\sig_{i|j})-\tau(\sig_{i|k})\right]\,.
\end{equation}
 For pure states, the minimum in \eq{taures} is always
attained by the decomposition realized with respect to the probe
mode $i$ with smallest local determinant
$\det{\sig_i}=m^2_{i|(jk)}$  \cite{contangle}.

\section{Distribution of Gaussian entanglement in de Sitter space \label{tools}}
\subsection{Generation of tripartite and bipartite  Gaussian entanglement}

As showed in Eq. (\ref{bogoliubov3}),  the Bunch-Davies vacuum for a  global  observer in de Sitter space can be expressed as a two-mode squeezed state of the $R$ and $L$ vacua \cite{Maldacena:2012xp,Kanno:2014lma}. Such state can be  obtained by
 $|0\rangle_{\rm BD}=\hat{U}_{R,L}(\gamma_p)|0\rangle_{R}|0\rangle_{L}$ in the Fock space, where
 $\hat{U}_{R,L}(\gamma_p)=e^{\gamma_p(\hat{c}^\dagger_{\text{R}}\hat{c}^\dagger_{\text{L}}-
\hat{c}_{\text{R}}\hat{c}_{\text{L}})}$ denotes the two-mode squeezing operator. In the phase space, the two-mode squeezing transformation  can be expressed by the symplectic  operator (for detail please see Appendix B)
\begin{eqnarray}\label{cmtwomode}
 S_{B,\bar B}(\gamma_p)= \frac{1}{\sqrt{1-|\gamma_p|^2}}
\left( {\begin{array}{*{20}{c}}
  I_2 & |\gamma_p| Z_2  \\
  |\gamma_p| Z_2& I_2 \\
\end{array}} \right),
\end{eqnarray}
where the basics  are   $|kl\rangle=|k\rangle_{B}|l\rangle_{\bar B}$, which denotes that the squeezing transformation is acting on the modes observed by Bob and anti-Bob ($\bar B$). In this paper, we assume that Alice is a  global observer who stays in the Bunch-Davies vacuum, while Bob is an observer resides  in the  region $R$ of the de Sitter open charts.
We further assume that,  there is no initial correlation between the entire state $\sigma^{\rm (G)}_{AB}(s)$ and the subsystem $\bar B$. Then the initial covariance matrix of the entire three mode state is  $\sigma^{\rm (G)}_{AB}(s) \oplus I_{\bar B}$, where  $\sigma^{\rm (G)}_{AB}(s)$
 is prepared by an entangled Gaussian two-mode squeezed state in  the Bunch-Davies vacuum
\begin{eqnarray}\label{inAR}
\sigma^{\rm (G)}_{AB}(s)=
\left( {\begin{array}{*{20}{c}}
   \cosh (2s)I_2 & \sinh (2s)Z_2  \\
   \sinh (2s)Z_2& \cosh (2s)I_2 \\
\end{array}} \right).
\end{eqnarray}
Here $I_2=\bigg( {\begin{array}{*{20}{c}}
   1 & 0  \\
 0& 1 \\
\end{array}} \bigg)$, $Z_2=\bigg(
                       \begin{array}{cc}
                         1 & 0 \\
                         0 & -1 \\
                       \end{array}
                    \bigg)$,
 and  $s$ is the squeezing parameter.
Under the transformation given in Eq. (\ref{cmtwomode}), the mode observed by Bob is mapped into two open charts. That is to say,
an extra set of modes $\bar{B}$  becomes relevant from the perspective of Bob in the open chart $R$. Therefore, a complete description of the three mode state after the curvature-induced  squeezing transformation  is
\begin{eqnarray}\label{in34}
\nonumber\sigma^{\rm (a)}_{AB \bar B}(s,\gamma_p) &=& \big[I_A \oplus  S_{B,\bar B}(\gamma_p)\big] \big[\sigma^{\rm (G)}_{AB}(s) \oplus I_{\bar B}\big]\\&& \big[I_A \oplus  S_{B,\bar B}(\gamma_p)\big]\,,
\end{eqnarray}
where $S_{B,\bar B}(\gamma_p)$ is the phase-space expression of the two-mode squeezing   in Eq. (\ref{cmtwomode}).

\begin{figure}[htbp]
\centering
\includegraphics[height=1.8in,width=2.0in]{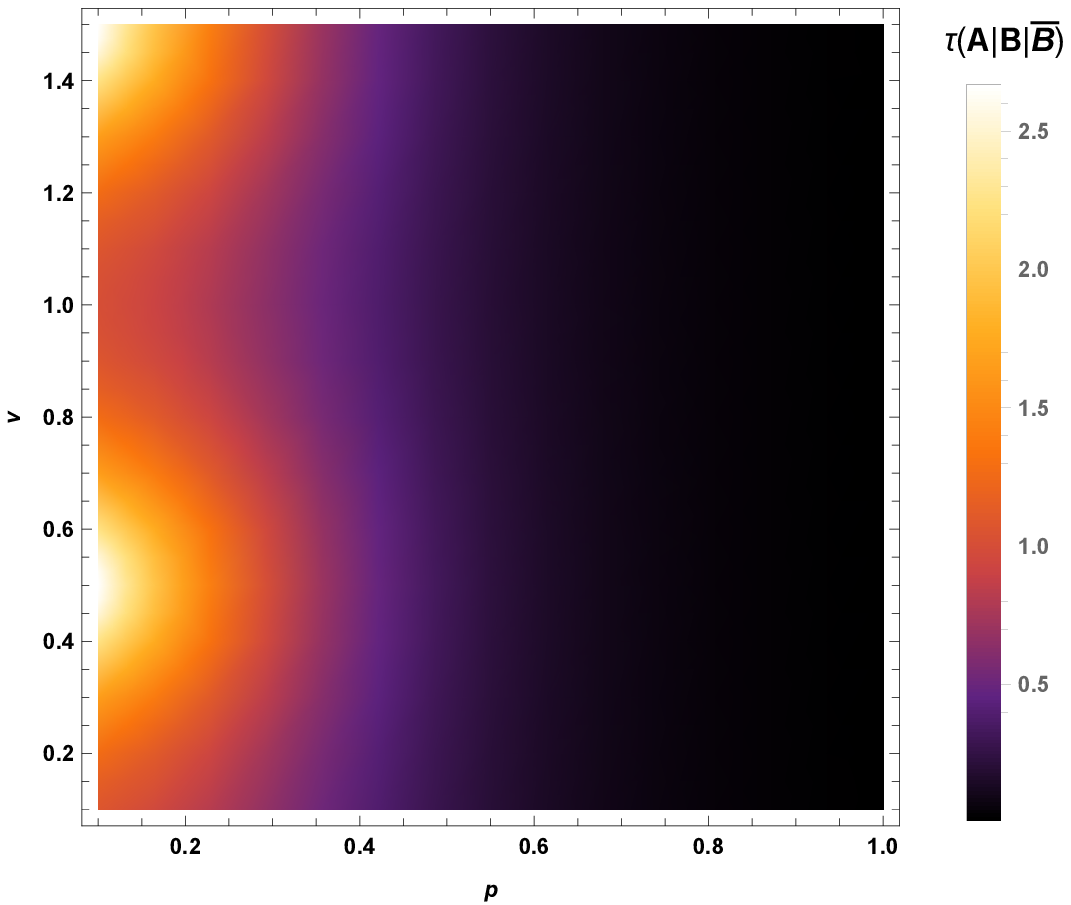}
\includegraphics[height=1.8in,width=2.0in]{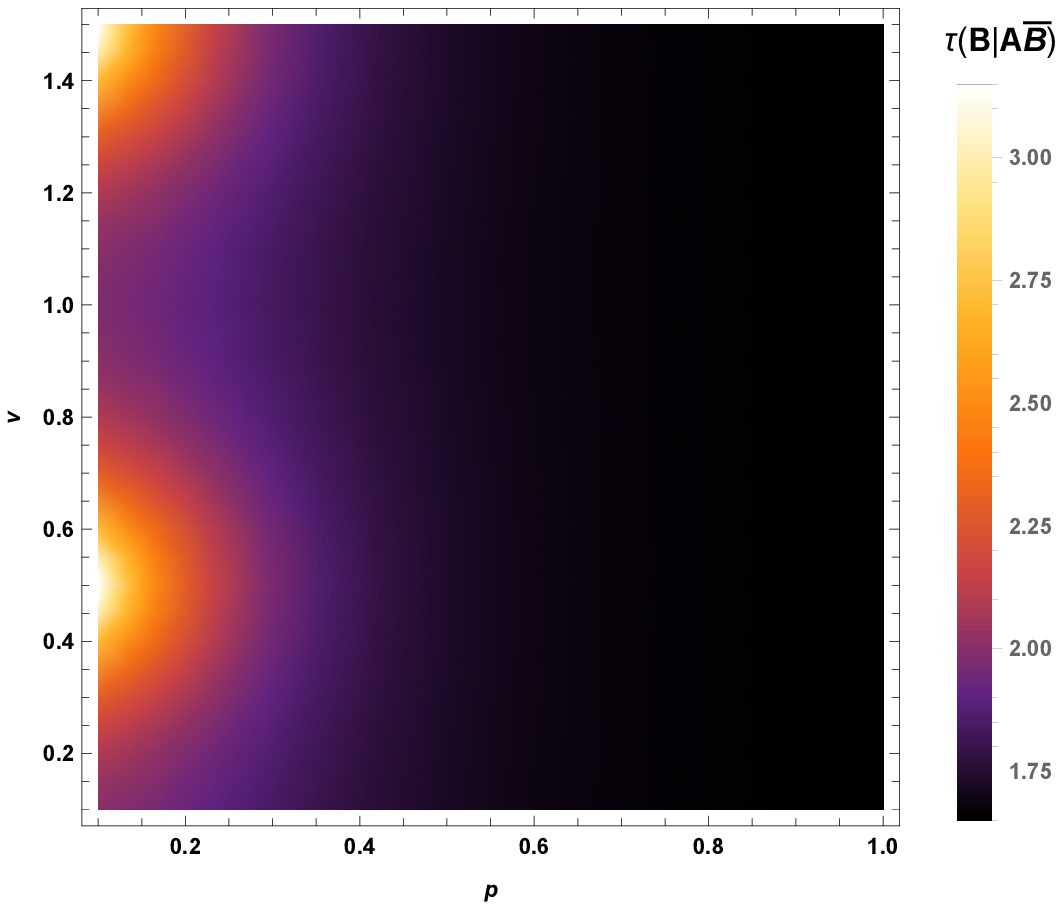}
\includegraphics[height=1.8in,width=2.0in]{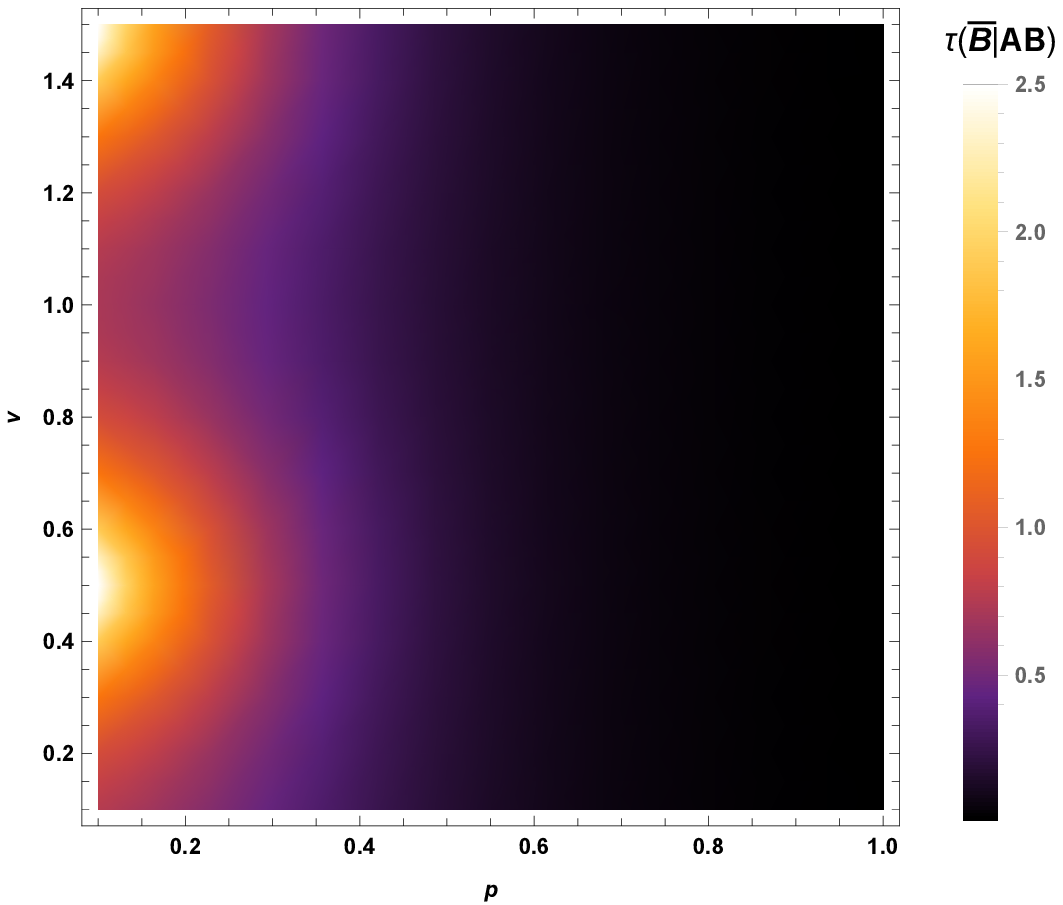}
\caption{ (Color online).  Gaussian quantum entanglement  as a function of the
 mass parameter $\nu$ and  curvature parameter of the de Sitter space $p$.  The squeezing parameter  $s$ in the initial state  is fixed as $s=1$.}\label{mmshow1}
\end{figure}

The  genuine  tripartite entanglement, as quantified by the residual contangle \eq{taures}, is found to be
\begin{eqnarray}
\tau_{(A|B|\bar B)} =4s^2- {\rm arcsinh}^2\bigg[ \sqrt{\frac{ |\gamma_p|^2 + (-2+|\gamma_p|^2
     ) \cosh (2 s)}{ \cosh (2 s) |\gamma_p|^2 + 2 -|\gamma_p|^2
              }-1}\bigg]\,. \label{m4_ar}
\end{eqnarray}
For any nonzero value of the squeezing parameters $s$ and $|\gamma_p|$, the residual contangle is nonzero.
This in fact indicates that
the state $\sig_{AB \bar B}$ is genuine  entangled
\cite{barbarella}: it contains genuine tripartite
entanglement among the global Alice, Bob in open chart $R$, and anti-Bob in
open chart $L$. In Fig. (\ref{mmshow1} a), we plot the genuine tripartite
entanglement $\tau_{(A|B|\bar B)} $ as a
function of the
 mass parameter $\nu$ and  curvature parameter of  de Sitter space $p$.  The initial prepared entanglement in Bunch-Davies vacuum is kept
fixed at $s=1.0$.

 From Fig. (\ref{mmshow1} a) we can see that the genuine tripartite entanglement monotonically decrease with  the increase of  curvature parameter $p$. We know that the curvature of de Sitter space is a descending function of $p$ because, as $p$ is closer to zero, the effect of  curvature gets stronger. Therefore,   this in fact indicates that space curvature in de Sitter space generates genuine tripartite entanglement  between the modes.   Very remarkably, the genuine tripartite increases with increasing acceleration $\nu$ when  $\nu<1/2$. It grows to a maximum for the conformally coupled
massless scalar field  ($\nu=1/2$) and then  decreases. Again, we get the maximum value of the genuine tripartite entanglement when the mass parameter  $\nu$  approaches the minimally coupled massless limit ( $\nu=3/2$).  This generation of quantum entanglement will be
precisely understood in the next subsection, where we will show that
the initial prepared entanglement does not disappear, but is
redistributed into tripartite correlations among different modes.

By using the definition Eq. (\ref{tau}), one can compute the bipartite contangles in different  $1 \to 2$ partitions of the state given in Eq. (\ref{in34}), which are found to be
\begin{eqnarray}\label{m3_12}
  \tau_{(A|B \bar B)} &=&{\rm arcsinh}^2\bigg[\sqrt{\cosh(2s)-1}\bigg]\,, \\
  \tau_{(B|A \bar B)} &=&{\rm arcsinh}^2\bigg[\sqrt{   \frac{\cosh (2 s)+2|\gamma_p|^2-1}{1-|\gamma_p|^2}
 }\bigg]\,, \nonumber \\
 \tau_{(\bar B|AB)} &=& {\rm arcsinh}^2\bigg[\sqrt{  \frac{|\gamma_p|^2(\cosh (2 s)+1)}{1-|\gamma_p|^2}
 }\bigg]\nonumber\,.
\end{eqnarray}
From Eq. (\ref{m3_12}) we can see that
 each single party is
entangled  with the block of the remaining two parties for any nonzero value of the  parameters $s$ and $\gamma_p$, with
respect to all possible global splitting of the three mode state. 

The values of the  $1 \to 2$ bipartite entanglement parameters from
\eq{m3_12} are plotted in Fig.~(\ref{mmshow1}) as a function of the
 mass parameter $\nu$ and curvature parameter  $p$, for a fixed degree of  squeezing parameter  $s=0.5$. From Fig. (\ref{mmshow1} b-c) we can see that the  $1 \to 2$ bipartite entanglement monotonically decrease with  the increase of  curvature parameter $p$, which means that space curvature also generates bipartite entanglement between these modes. However, the generated entanglement are  apparently affected by the  curvature of de Sitter space only around $\nu=1/2 $ (conformal scalar limit) and $\nu=3/2$ (massless scalar limit).   The degree of entanglement  is much bigger than others  when the value of $\nu$ is in the neighborhood of $\nu=1/2 $ and $\nu=3/2$, as clearly visible in  Fig. (1 b-c).

 \begin{figure}[htbp]
\centering
\includegraphics[height=1.8in,width=2.0in]{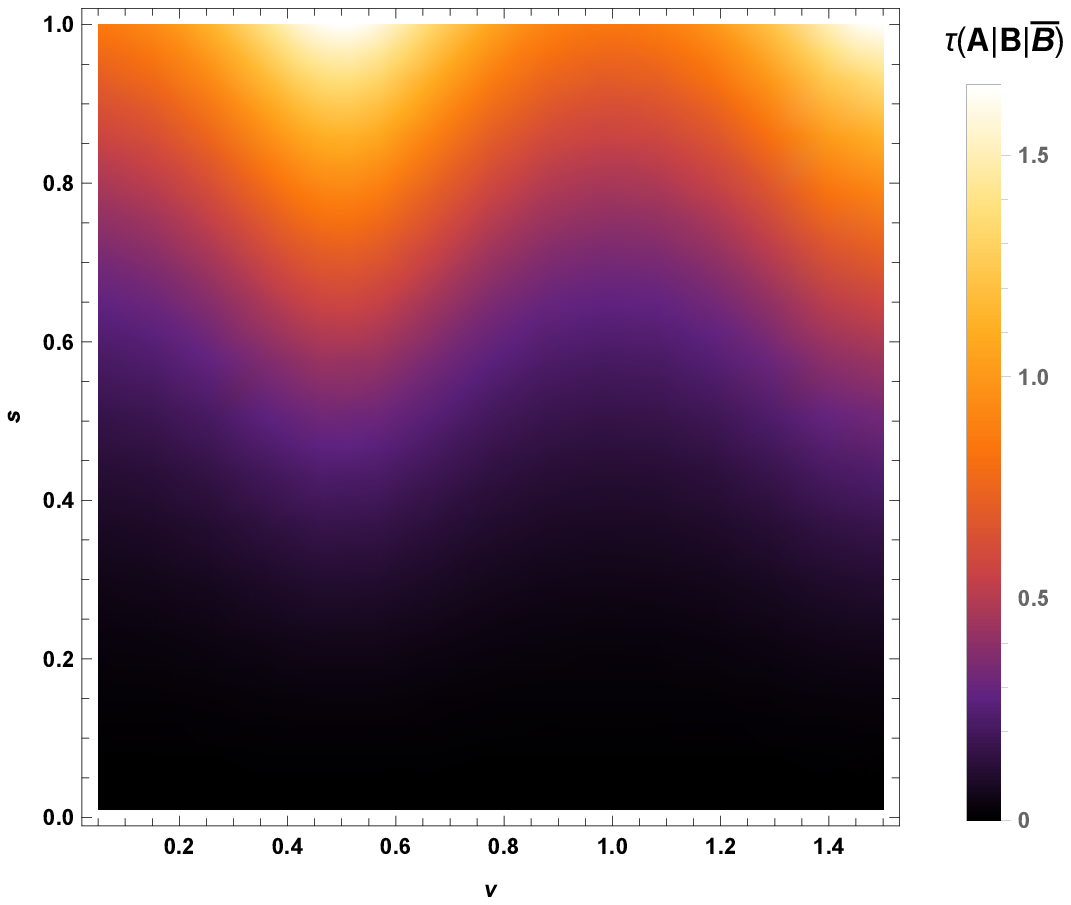}
\includegraphics[height=1.8in,width=2.0in]{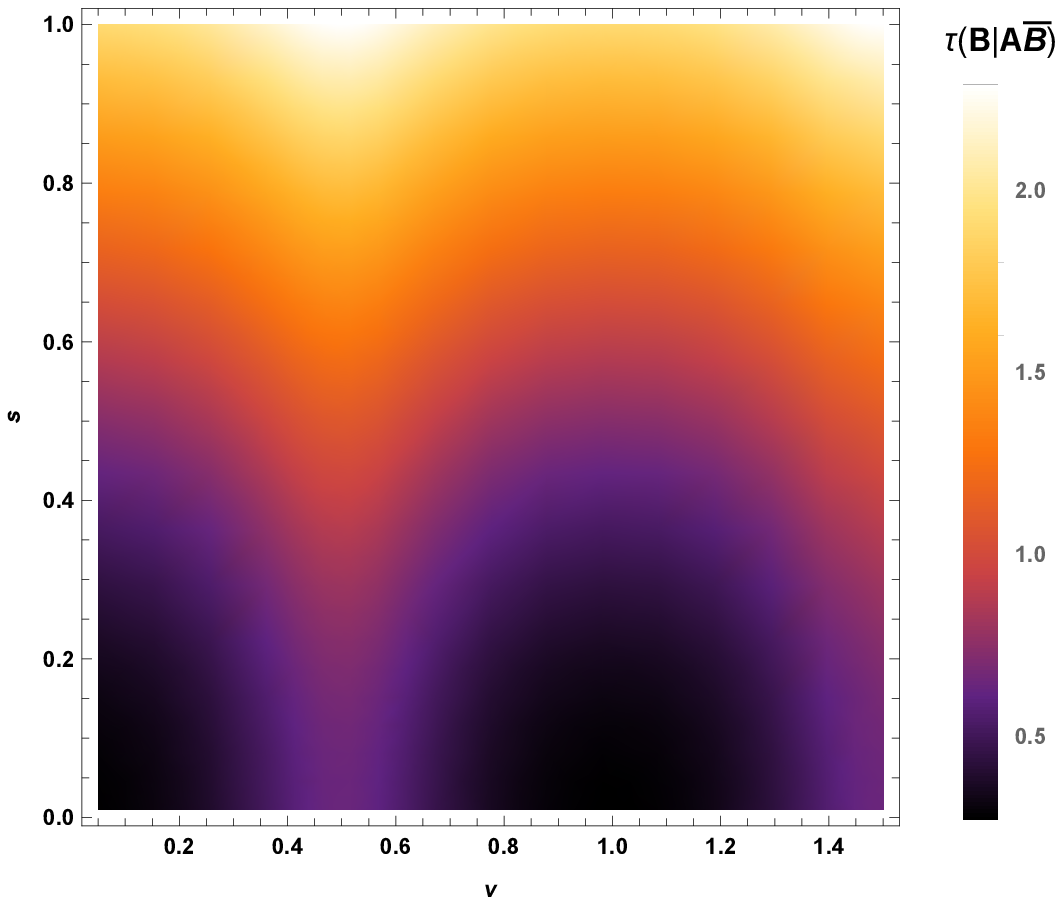}
\includegraphics[height=1.8in,width=2.0in]{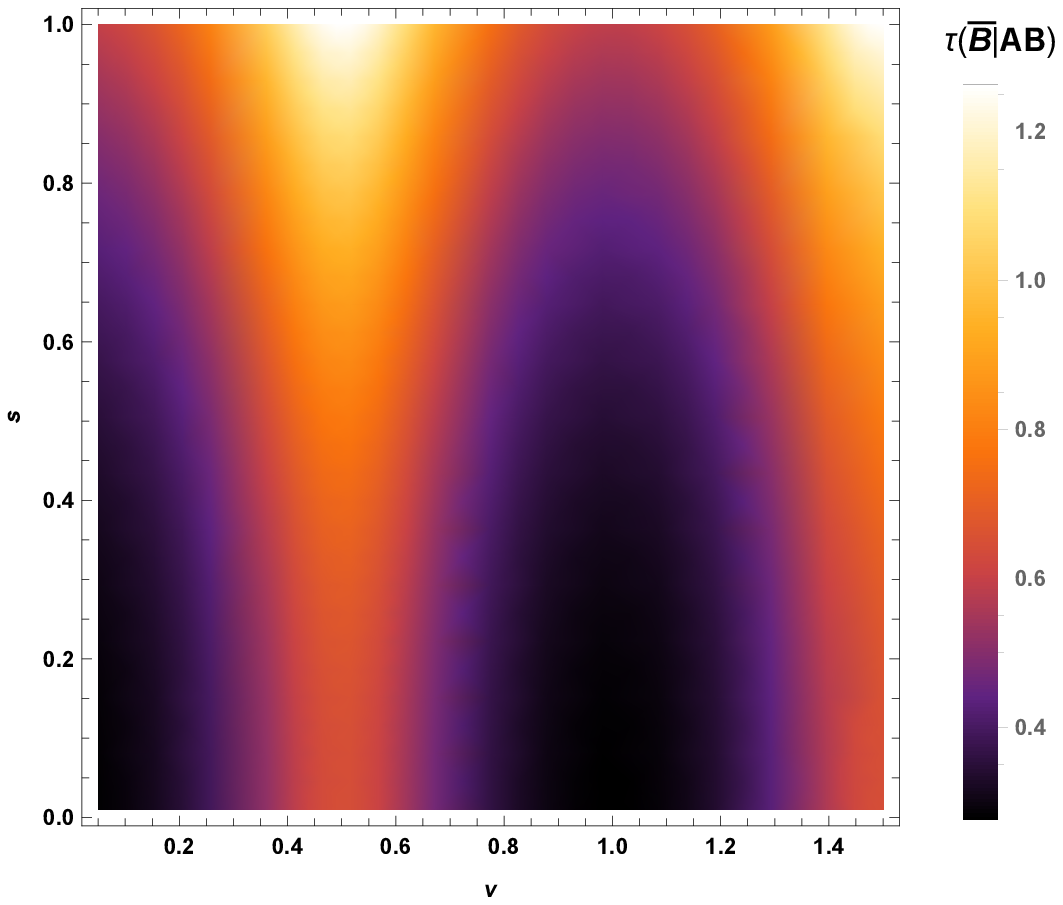}
\caption{ (Color online).  Gaussian quantum entanglement   as a function of  the squeezing parameter $s$
and the mass parameter $\nu$.  The curvature parameter of the de Sitter space is fixed as $p=0.2$.}\label{mmshow2}
\end{figure}

In Fig. (\ref{mmshow2} a), we plot the genuine tripartite
entanglement $\tau_{(A|B|\bar B)} $ as a
function of  the squeezing $s$ and
 mass parameter $\nu$ with fixed $p=0.2$.
 The bipartite quantum entanglement  between
the mode described by one observer and the group of modes described
by the other two is plotted in Fig.~(\ref{mmshow2} b-c). From (\ref{mmshow2} a-c) we can see that both the generated tripartite and bipartite are monotonous increasing function of
 the  initial  squeezing parameter $s$ in the Bunch-Davies vacuum. This confirms the fact that the distributed  quantum resource in these mode root in the prepared resource in the two-mode squeezed state in Eq. (\ref{inAR}). Due to
the monogamy of entanglement, the entanglement between the initial modes
is degraded. In addition, comparing to the bipartite contangles, the genuine tripartite
entanglement $\tau_{(A|B|\bar B)}$ is found to be less sensitive to the mass parameter $\nu$.

\subsection{Sharing of  Gaussian entanglement among  causally disconnected regions}

To better understand  the interplay
between squeezing and the space curvature in the generation of
 Gaussian quantum entanglement,  in this subsection we  discuss the behavior of  $1\to 1$ bipartite entanglement in  the tripartite system. First of all, we find that there is no bipartite entanglement between the modes
observed by the globe Alice and anti-Bob in the open chart $L$. Then we discuss the entanglement between the initial nonseparable Alice and Bob and the initial separable Bob and anti-Bob.  Because Bob in  chart $R$ has no access to the modes in the causally disconnected $L$ region, we must therefore trace over the inaccessible modes.  Taking the
trace over mode $\bar B$ in chart $L$,  one obtains
covariance matrix $\sigma_{AB}(s,\gamma_p)$ for Alice and Bob
\begin{eqnarray}\label{CM1}
\sigma_{AB}(s,\gamma_p)=\left( {\begin{array}{*{20}{c}}
 \cosh(2s) I_2 & \frac{\sinh(2s)}{\sqrt{1-|\gamma_p|^2}} Z_2  \\
  \frac{\sinh(2s)}{\sqrt{1-|\gamma_p|^2}} Z_2& \frac{|\gamma_p|^2+\cosh(2s)}{1-|\gamma_p|^2}I_2 \\
\end{array}} \right)
.
\end{eqnarray}

 We also interested in the  Gaussian entanglement between mode $B$ in the $R$ region  and anti-Bob in $L$ region, which are  separated by the event horizon of the de Sitter space.
Tracing over the modes in $A$, we obtain the covariance matrix
$\sigma_{B\bar B}(s,\gamma_p)$ for Bob and anti-Bob

\begin{eqnarray}\label{CM22}
\sigma_{B\bar B}(s,\gamma_p)=\left( {\begin{array}{*{20}{c}}
  \frac{|\gamma_p|^2+\cosh(2s)}{1-|\gamma_p|^2} I_2 & \frac{2|\gamma_p|\cosh^2(s)}{1-|\gamma_p|^2} Z_2  \\
  \frac{2|\gamma_p|\cosh^2(s)}{1-|\gamma_p|^2} Z_2& \frac{1+|\gamma_p|^2\cosh(2s)}{1-|\gamma_p|^2}I_2 \\
\end{array}} \right).
\end{eqnarray}

Then we find that the modes observed by Bob and anti-Bob are entangled when the curvature is strong enough  even though they are separated by the event horizon,  which verifies the fact that the  entanglement is one kind of  nonlocal quantum correlation.

\begin{figure}[htbp]
\centering
\includegraphics[height=2.0in,width=3.9in]{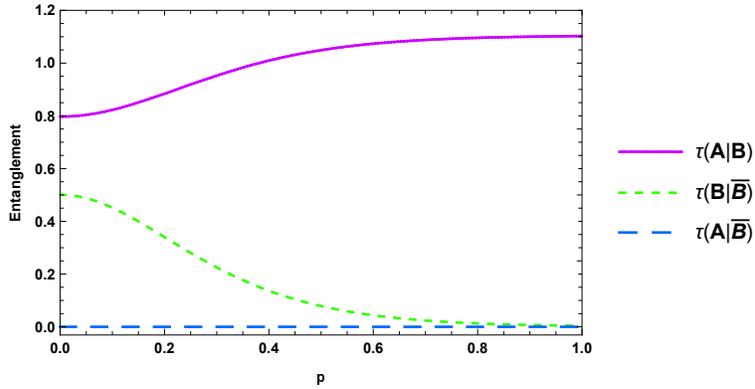}
\caption{ (Color online). Bipartite entanglement
between the modes as a function of the  curvature parameter of the de Sitter space $p$. The initial  squeezing parameter is fixed as $s=0.8$ and the
 mass parameter is fixed as  $\nu=0.2$.}\label{Fig4}
\end{figure}

The contangles of the  states $\sigma_{AB}$ and $\sigma_{B\bar B}$, quantifying the bipartite
entanglement described by two  observers, are found to be
\begin{eqnarray}
\tau_{(A|B)} &=& {\rm arcsinh}^2\bigg[\frac{ |\gamma_p|^2 + (2-|\gamma_p|^2
     ) \cosh (2 s)}{ \cosh (2 s) |\gamma_p|^2 + 2 -|\gamma_p|^2
              }\bigg]\,, \label{m3_ar} \\
              \tau_{(B|\bar B)} &=& {\rm arcsinh}^2\bigg( \frac{1+|\gamma_p|^2}{1-|\gamma_p|^2}\bigg) \label{m3_rr}\,.
\end{eqnarray}

 Let us give some comments on the quantum entanglement created between the
modes in the open charts $R$ and $L$. Note that the
entanglement in the mixed state $\sig_{B \bar B}$ is exactly equal
 to that of a pure two-mode squeezed state with squeezing
$\gamma_p$. This provides a clearcut interpretation of the two-mode squeezing
mechanism, in which the curvature of de Sitter space is responsible of the
creation of entanglement between the accessible mode
described by Bob, and the unaccessible one described by anti-Bob.

From Fig. (\ref{Fig4}) we can see that the  bipartite entanglement between the initially entangled Alice and Bob monotonically increases with  increasing $p$, while the entanglement between the causally disconnected Bob and anti-Bob  decreases with  increasing $p$.
Now we have all the elements necessary to fully understand the spacetime curvature on  Gaussian entanglement of scalar fields in the de Sitter space: There is bipartite entanglement between the two modes in the two distinct open charts, and this entanglement is a function of the spacetime curvature parameter.
In addition,  the bipartite entanglement initially prepared in the Bunch-Davies vacuum is
redistributed into a genuine tripartite entanglement among the modes
described by the observers in different open charts. Therefore, as a consequence of
the monogamy of entanglement, the entanglement between the two modes
 $A$ and $B$ is degraded.

\section{Conclusions}

 We have studied the distribution  of entanglement  among the mode $A$ described by the globe  observer Alice, mode $ B$ described by Bob in the de Sitter region $R$, and the complimentary mode $\bar B$ described by a hypothetical observer anti-Bob in  the  causally disconnected region $L$. It is found  that space curvature in de Sitter space generates genuine tripartite entanglement  between the modes. In the expanding de Sitter space,
 each single party in the tripartite quantum system is
entangled  with the block of the remaining two parties, with
respect to all possible global splitting of the three mode state.  The modes $B$ and $\bar B$ get entangled when the curvature is strong enough  even though they are separated by the event horizon. In addition, comparing to the generated bipartite entanglement, the  generated genuine tripartite
entanglement  is found to be less sensitive to the mass parameter $\nu$.  The modes observed by Bob and anti-Bob are entangled when the curvature is strong enough  even though they are separated by the event horizon.  This provides an  interpretation of the two-mode squeezing
mechanism, in which the curvature of de Sitter space is responsible of the
creation of entanglement between them.  We also find that the effects of the  curvature of de Sitter space on the generated tripartite and bipartite entanglement become more apparent  in the limit of  conformal and massless scalar fields.

\begin{acknowledgments}
This work is supported by the National Natural Science Foundation
of China under Grant  No. 11675052; and the  Natural Science Fund  of Hunan Province  under Grant No. 2018JJ1016; and Science and Technology Planning Project of Hunan Province under Grant No. 2018RS3061.	

\end{acknowledgments}

\appendix
\onecolumngrid

\section{The definition of contangle for   continuous variables}
In this appendix, we recall the definition of contangle for  continuous variables and  how it relates logarithmic negativity. Let us start with the  monogamy inequality for a three mode
Gaussian state
\begin{equation}\label{CKWine}
E^{i|(jk)}- E^{i|j} - E^{i|k} \ge 0 \; ,
\end{equation}
where $E$ is a proper measure of bipartite  entanglement and the
indexes $\{i,j,k\}$ label the modes.
When dealing with
$1\times N$ partitions  of a multimode pure Gaussian
state together with its $1 \times 1$ reduced partitions, the
measure of entanglement should be a monotonically decreasing function $f(\tilde n_-)$ of
the smallest symplectic eigenvalue $\tilde n_-$ of the corresponding
partially transposed covariance matrix $\tilde{\sigma}$. This is because
 $\tilde n_-$ is the only eigenvalue that can be
smaller than $1$ \cite{adescaling}, which violates  the PPT criterion for  the selected bipartition. Moreover, for a pure
three mode Gaussian state, it is required that the
bipartite entanglements $E^{i|(jk)}$ and $E^{i|j}=E^{i|k}$ are
respectively functions $f(\tilde n_-^{i|(jk)})$ and $f(\tilde
n_-^{i|j})$ of the associated smallest symplectic eigenvalues
$\tilde n_-^{i|(jk)}$ and $\tilde n_-^{i|j}$ \cite{contangle}. 

For a generic pure state
$|\psi\rangle$ of a $(1+N)$ mode continuous variable system, one can define  the square of
the logarithmic negativity as a  measure of bipartite entanglement:
\begin{equation}
\label{etaupure}\tau (\psi) \equiv \ln^2 \| \tilde \rho \|_1 \; ,
\quad \rho = \vert\psi \rangle \! \langle \psi \vert\; .
\end{equation}
This is  a convex,
increasing function of the logarithmic negativity $E_\N$.
For any pure multipartite Gaussian state $|\psi \rangle$ with covariance matrix $\sigma$,
explicit evaluation gives
\begin{equation}
\label{piupurezzapertutti}\tau (\psi) \equiv \tau (\sigma) = \ln^2 \left(1/\mu_A -
\sqrt{1/\mu_A^2-1}\right),
\end{equation}
where $\mu_A = 1/\sqrt{\det\sigma_A}$ is the local purity of the reduced
 covariance matrix $\sigma_A$. From Eq. (\ref{piupurezzapertutti})  we can see
that the convex roof of the squared logarithmic negativity defines
the continuous-variable tangle, or, in short, the
{\it contangle} for pure states \cite{contangle}.
The definition in Eq.~(\ref{etaupure}) can be naturally extended
to mixed states $\rho$ of $(N+1)$ mode continuous variable systems through
the convex-roof formalism \cite{contangle}.
Namely, we can define the  contangle $\tau(\rho)$  for mixed states as
\begin{equation}\label{etaumix}
\tau(\rho) \equiv \inf_{\{p_i,\psi_i\}} \sum_i p_i
\tau(\psi_i)\; ,
\end{equation}
where the infimum is taken over all convex decompositions of $\rho$
in terms of pure states $\{|\psi_i\rangle\}$.

The sum in \eq{etaumix} should be replaced by an integral if the index $i$ is
continuous, and
the probabilities $\{p_i\}$ is replaced by the probability distribution
$\pi(\sigma)$.   For mixed multimode Gaussian states with  covariance matrix $\sigma$, one should
denote the contangle by $\tau(\sigma)$, in analogy with the notation
used for pure Gaussian states.  Then we  define the  contangle for Gaussian states by the infimum of the average contangle, taken
over all pure Gaussian state,
\begin{equation}
\tau(\sigma) \equiv \inf_{\{\pi(d\sigma ), \sigma \}}
\int \pi (d\sigma)\tau (\sigma) \; .
\label{GaCoRo}
\end{equation}
If $\sigma_{AB}$ denotes a mixed
two-mode Gaussian state,  the Gaussian decomposition is the
optimal one \cite{contangle}, and the
optimal pure state covariance matrix $\sigma_{AB}$ minimizing $\tau(\sigma_{AB})$ is
characterized by $\tilde
n_-(\tilde{\sigma_{AB}})$. Considering that the smallest
symplectic eigenvalue is the same for both partially transposed covariance matrixes, we have
 $\tau(\sigma_{AB})  = [\max\{0,-\ln \tilde
n_-(\sigma_{AB})\}]^2$. Therefore, for a mixed bipartite Gaussian state where
subsystem $A$ comprises one mode only,  the contangle $\tau$ can
be computed by  \cite{contangle}
\begin{equation}
 \tau (\sigma_{AB} )\equiv \tau (\sigma_{A\vert
B}^{opt} )=g[m_{AB}^2 ],\;\;\;g[x]={\rm arcsinh}^2[\sqrt {x-1}],
\end{equation}
where $\sigma_{AB}^{opt} $ corresponds to a pure Gaussian state,
and $m_{AB} \equiv m(\sigma_{AB}^{opt} )=\sqrt {\det
\sigma_A^{opt} } =\sqrt {\det \sigma_B^{opt} } $.

\section{Two-mode squeezed transformation  in phase space and the  entire final state}

 The Bunch-Davies vacuum for a  global  observer can be expressed as a two-mode squeezed state of the $R$ and $L$ vacua
\begin{eqnarray}
|0\rangle_{\rm BD}=\sqrt{1-|\gamma_p|^2}\,\sum_{n=0}^\infty\gamma_p^n|n\rangle_L|n\rangle_R\,,
\label{bogoliubov2}
\end{eqnarray}
where $\gamma_p$ is the  squeezing parameter.  In the Fock space, the  two-mode squeezed state can be obtained by
 $|0\rangle_{\rm BD}=\hat{U}_{R,L}(\gamma_p)|0\rangle_{R}|0\rangle_{L}$, where
 $\hat{U}_{R,L}(\gamma_p)=e^{\gamma_p(\hat{c}^\dagger_{\text{R}}\hat{c}^\dagger_{\text{L}}-
\hat{c}_{\text{R}}\hat{c}_{\text{L}})}$  is the two mode squeezing operator. In the phase space, such transformation  can be expressed by a symplectic phase-space operator 
\begin{equation}\label{cmtwomo1}
\sigma_{B\bar B}(\gamma_p)=
\begin{pmatrix}
\cosh \left(2\gamma_p\right) & 0 & \sinh \left(2\gamma_p\right)
& \sinh\left( 2\gamma_p\right)
 \\
0 & \cosh \left(2\gamma_p\right) & \sinh \left(2\gamma_p\right)
& -\sinh\left( 2\gamma_p\right)  \\
\sinh \left(2\gamma_p\right)
& \sinh \left(2\gamma_p\right)
& \cosh\left(2\gamma_p\right) & 0 \\
\sinh\left( 2\gamma_p\right)
& -\sinh\left( 2\gamma_p\right)
& 0 & \cosh\left(2\gamma_p\right)
\end{pmatrix},
\end{equation}
where $\cosh \gamma_p=(\sqrt{1-|\gamma_p|^2})^{-1}$.
This  covariance matrix is computed by $\sigma_{B\bar B}(\gamma_p)= S^T_{B,\bar B}(\gamma_p)I_4 S_{B,\bar B}(\gamma_p)$, where (Eq. 15 in the main manuscript)
\begin{eqnarray}\label{cmtwomode1}
 S_{B,\bar B}(\gamma_p)= \frac{1}{\sqrt{1-|\gamma_p|^2}}\left(\!\!\begin{array}{cccc}
1&0&|\gamma_p|&0\\
0&1&0&-|\gamma_p|\\
|\gamma_p|&0&1&0\\
0&-|\gamma_p|&0&1
\end{array}\!\!\right),
\end{eqnarray}
which denotes that squeezing transformation are performed to the bipartite state shared between Bob and anti-Bob ($\bar B$).

The expression given in  Eq. (17) of  the main manuscript is the phase space  description of the entire state after the curvature-induced  squeezing transformation given in Eq. (15). The covariance matrix of the  entire state is computed by
\begin{eqnarray}
\nonumber\sigma^{\rm }_{AB \bar B}(s,r) &=& \big[I_A \oplus  S_{B,\bar B}(\gamma_p)\big] \big[\sigma^{\rm (M)}_{AB}(s) \oplus I_{\bar B}\big]\\&& \nonumber\big[I_A \oplus  S_{B,\bar B}(\gamma_p)\big]\\
 &=& \left(
       \begin{array}{ccc}
          \mathcal{\sigma}_{A} & \mathcal{E}_{AB} & \mathcal{E}_{A\bar B} \\
         \mathcal{E}^{\sf T}_{AB} &  \mathcal{\sigma}_{B} & \mathcal{E}_{B\bar B} \\
         \mathcal{E}^{\sf T}_{A\bar B} & \mathcal{E}^{\sf T}_{B\bar B} &  \mathcal{\sigma}_{\bar B} \\
       \end{array}
     \right)
 \,,
\end{eqnarray}
where $\sigma^{\rm (G)}_{AB}(s) \oplus I_{\bar B}$ is the initial covariance matrix for the entire system.
In Eq. (\ref{in34}) the diagonal elements have the following forms:  \begin{equation} \mathcal{\sigma}_{A}=\cosh(2s)I_2,\end{equation} \begin{equation}\mathcal{\sigma}_{B}=[\cosh(2s) \cosh^2(\gamma_p) + \sinh^2(\gamma_p)]I_2,\end{equation} and \begin{equation}\mathcal{\sigma}_{\bar B}=[\cosh^2(\gamma_p) + \cosh(2s) \sinh^2(\gamma_p)]I_2. \end{equation} The non-diagonal elements are 
$\mathcal{E}_{AB}=[\cosh(\gamma_p) \sinh(2s)]Z_2$, \begin{equation}\mathcal{E}_{B\bar B}=[\cosh^2(s) \sinh(2\gamma_p)]Z_2,\end{equation} and \begin{equation}\mathcal{E}_{A\bar B}=[\sinh(2s) \sinh(\gamma_p)]Z_2, \end{equation} with
\begin{equation}
\label{eq:def1}
Z_2=
\begin{pmatrix}
1 & 0 \\
0 & -1
\end{pmatrix}.
\end{equation}

\end{document}